\documentclass[12pt,twocolumn]{article}
\usepackage[a4paper,left=20mm,right=20mm,top=25mm,bottom=25mm,includeheadfoot]{geometry}

\usepackage{mathpazo}
\usepackage{graphicx}
\usepackage{amsmath,textcomp}
\usepackage{multirow}
\usepackage{makeidx}
\makeindex

\usepackage{sectsty}
	\allsectionsfont{\sffamily\raggedright}
	\sectionfont{\sffamily\large\raggedright}
	\subsectionfont{\sffamily\normalsize\raggedright}
\usepackage{ftnright}

\usepackage{flushend}
\usepackage{cuted}
\usepackage{color}
\usepackage{graphicx}
\usepackage{hyperref}
\usepackage{pstricks,amssymb}

\usepackage{fancyhdr}
\begin{document}
\title{\vspace{-2em}\bfseries\sffamily Horizon, homogeneity and flatness problems -- do their resolutions really depend upon inflation?}
\author{\normalsize Ashok K. Singal\\
Astronomy and Astrophysics Division, Physical Research Laboratory\\
Navrangpura, Ahmedabad 380 009, India.\\
{\tt ashokkumar.singal@gmail.com}
}
\date{\itshape Submitted on xx-xxx-xxxx}
\maketitle
\thispagestyle{fancy}

\begin{abstract}
{\sffamily
We point out that the horizon problem encountered in standard text-books or review papers on cosmology is, in general, derived for world models based on Robertson-Walker line element where homogeneity and isotropy of the universe -- \`a la cosmological principle -- is assumed to begin with and is guaranteed for all epochs. Actually what all happens in that scenario is that in such a universe, whose  evolutionary behaviour is described by a single scale factor, which may be time dependent but is  otherwise independent of spatial coordinates, the light signals in a finite time might not be covering all the available space. Further, the flatness problem, as it is posed, is not even falsifiable. The usual argument offered in the literature is that the present density of the universe is very close to the critical density value and that the universe must be flat since otherwise in past at $\sim10^{-35}$ second (near the epoch of inflation) there will be extremely low departures of density from the critical density value (of the order $\sim10^{-53}$), requiring a sort of fine tuning. We show that even if the present value of the density parameter were very different, still at $10^{-35}$ second it would differ from unity by the same fraction.  Thus a use of fine tuning argument to promote $k = 0$ model amounts to {\em a priori} rejection of all models with $k \ne 0$. Without casting any aspersions on the inflationary theory, which after all is the most promising paradigm to explain the pattern of anisotropies observed in the cosmic microwave background, we argue that one cannot use homogeneity and flatness in support of inflation.
}\\ 
\hrule
\end{abstract}

\section{Introduction}
In textbooks and review articles on modern cosmology \cite{29,10a,1,9,10,L08,6} one almost invariably comes across a section devoted to the subject of 
observed homogeneity and near-flatness of the universe where it is argued that to explain these observations inflation is almost 
a must. In fact that was the prime motive of Guth \cite{2} to propose inflation in the first place. 
We show that the arguments offered therein are not proper. The horizon problem, which leads to the causality arguments for the  homogeneity, is derived in the Friedmann-Robertson-Walker (FRW) world models where homogeneity and isotropy of the universe at some large enough scale, i.e. Cosmological Principle (CP), is presumed to begin with. We have no idea whether a horizon problem would still 
arise in non-homogeneous world models that do not depend upon the Robertson-Walker line element. Therefore as long as we confine ourselves to investigating properties of FRW world models, there is no homogeneity issue. 

To justify flatness,  the usual argument employed in the literature is that 
the present density of the universe is very close (within an order of magnitude) to the critical density value. From this one infers 
that the universe must be flat since otherwise in past at $10^{-35}$ second (near the epoch of inflation) there will be extremely low departures  
of density from the critical density value (i.e., differing from unity by a fraction of order $\sim 10^{-53}$), requiring a sort of fine tuning. 
Actually we show that even if the present value of the density parameter (in terms of the critical density value) were very different, still 
at $10^{-35}$ second it would in any case differ from unity by a fraction of order $\sim 10^{-53}$. 
Therefore such a fine-tuning does 
not discriminate between various world models and a use of fine tuning 
argument amounts to  {\em a priori} rejection of all models with $k \ne 0$, because inflation or no inflation, the density parameter in all 
FRW world models gets arbitrarily close to unity as we approach the epoch of the big bang. Thus the flatness problem, as posed in literature, is not even falsifiable,
as that way, without even bothering to measure the actual density, 
we could use any sufficiently early epoch and use ``extreme fine-tuning'' arguments to rule out all non-flat models. Thus without casting 
any whatsoever aspersions on the inflationary theories, we point out that one cannot use these type of arguments, viz. homogeneity and 
flatness, in support of inflation.

It should be pointed out that inflation is not the only scenario proposed to help resolve the horizon problem. Bouncing scenarios \cite{Br12, Li15} have been proposed as an alternative to inflation. Apart from these, the loitering universe models have also been suggested to overcome the horizon problem \cite{Sa92, El05}. However, amongst these various alternatives, inflation seem to be providing the simplest mechanism to overcome the horizon problem and reduce the extent of fine tuning required to explain the observed levels of spatial flatness. In any case, as would become apparent, our critique is as much applicable to these alternate theories as to inflation.

It is also necessary to emphasize here that we are not casting 
any, whatsoever, aspersions on the inflationary theory itself, which after all is the most promising paradigm to generate the nearly scale invariant spectra of primordial perturbations that seems to be required to explain the pattern of anisotropies observed in the CMBR. Moreover, there is no doubt about the robustness of inflation to inhomogeneous initial conditions \cite{Cl17, Au20}. In fact, the motivation for inflation may instead lie in the microphysical explanation of spatially flat FRW cosmology, i.e. from the point of view of high energy physics, why do we find ourselves in a spatially flat FRW spacetime? Cosmic inflation is a mechanism to create macroscopically large regions of spacetime which are described by spatially flat FRW metric, in a microscopic theory which operates at very small length scales. 

As we shall spell out, our contention here is only that one cannot use the arguments based on  homogeneity and flatness, initially  offered \cite{2} and subsequently repeated in most textbooks and review articles \cite{29,10a,1,9,10,L08,6}, in support of inflation.

\section{Horizon and Homogeneity Problem}
A particle horizon in the cosmological context implies a maximum distance yonder which we as observers have not yet seen the universe due perhaps to a 
finite speed of light as well as a finite age of the universe. In other words these are the farthest regions of the universe 
(redshift $z \rightarrow \infty$) from which the light signals have just reached us. However when we look at the universe we find 
that distant regions in opposite directions seen by us have similar cosmic microwave background radiation (CMBR) temperatures.
The particle
horizon problem in standard cosmological big bang model is that these different regions of the universe have 
not ever communicated with each other, but nevertheless they seem to have the same temperature, 
as shown by the CMBR which shows almost a uniform temperature ($2.73\:^\circ$K) across the sky, irrespective of the direction.
How can this be possible, considering that any exchange of information (say, through photons or any other means) can occur, at most, at 
the speed of light. 
How can such two causally disconnected regions have one and same temperature, unless one makes a somewhat ``contrived'' presumption that 
the universe was homogeneous and isotropic to begin with when it came into existence (see e.g. \cite{1}).
\begin{figure}
\begin{center}
\includegraphics[width=8cm]{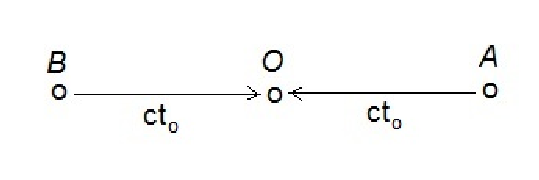}
\caption{An observer at $O$ (us) receiving signal from distant objects at $A$ and $B$ at time $t_{\rm o}$, which is the time since 
big bang. Signal from $A$ having just reached $O$ could not have yet reached $B$ and vice versa. In practice, the farthest that we can observe is the CMBR from $A$ and $B$, then $t_{\rm o}$ is the time since the recombination era when radiation and matter got decoupled.} 
\end{center}
\end{figure}

One can illustrate the particle
horizon problem using a simple, though somewhat naive, argument in the following way (more sophisticated treatment using conformal diagrams can be found in literature, see e.g. \cite{22}). 
According to the big bang model the universe has only a finite age, say $t_{\rm o}$. Then light (or information) from regions at a cosmological 
distance $ct_{\rm o}$ from us would have reached us just now, and could not have crossed over to similar distant regions on the other side of 
us. Then how come two far-off regions on two opposite sides of us have managed to achieve the 
homogeneity so that we see them having same properties? Though this simple argument does contain 
an element of truth, but it could not be always true and its naive nature can be seen from the simplest of FRW models, namely empty 
universe of Milne ($\rho = 0$), where 
the worldline of even the most distant object in the universe is, up
to a certain instant, within horizon, let it be in any region in any direction from us in this infinite 
universe model. For instance, in this world model, an observer at $B$, at any time, will receive signals from $O$ as well as $A$ (assuming both to be at $z \rightarrow \infty$) simultaneously. In fact all 
regions in this universe at any time receive past signals from all other regions even though the universe is infinite (see also \cite{22}). 
Thus horizon problem does not arise in this particular world model. However, in more realistic cases of general relativistic cosmological 
models, say with finite density, almost invariably one comes across horizon problems. 

From the observed CMBR, the universe appears to be very close to isotropic. At the same time 
Copernican principle states that earth does not have any eminent or privileged position in the universe and therefore an observer's 
choice of origin should have no bearing on the appearance of the distant universe. 
From this we infer that the cosmos should appear isotropic from any vantage point in the universe, which directly implies homogeneity. 
Then one can use Weyl's postulate of an infinite set of equivalent {\em fundamental observers} (FOs) spread around the universe, 
who agree on a ``global'' time parameter, orthogonal to 3-d space-like hyper surfaces \cite{1,4,5,Ho06}, and measured using some local observable like density, 
temperature, pressure etc. as a parameter.
Thus holding CP to be true all FOs, one gets for such observers a metric for the universe known as Robertson-Walker metric. It might be emphasized that the underlying assumption of CP is all pervasive in all modern cosmological models of the universe \cite{Ma11,Cl12}. This is because it is CP alone that allows Weyl's postulate of a set of equivalent FOs in the universe, who agree on a ``global'' time parameter, which in turn allows a single scale factor R(t) applicable to all parts of the universe.

Now in any cosmological scenario with an initial singularity (big bang), at least in all models where inflation theory is applied, a single scale factor R(t) is used, even for the exponential expansion in the inflation era, and that implicitly assumes CP. And when we explore these models, we encounter horizon where light signal between different FOs may not have yet got exchanged (particle
horizon) or may never get exchanged (event horizon), but that does not alter the implicit assumption of CP for all of them. From that we can readily infer that even if there may be a horizon problem, it does not necessarily imply that we have a homogeneity problem too. In essence, in all cosmological models where inflation is being applied, to purportedly alleviate the homogeneity problem, we have a priory assumption that CP holds good. 

Is there any other evidence in support of the CP? Optically the universe shows structures up to the scale of 
super clusters of galaxies and even up to hundreds of mega parsecs \cite{Park17}, although still beyond it might be more uniform \cite{Nt17,Go18}. The conventional wisdom is that when observed on a
sufficiently large scale the universe would appear homogeneous and isotropic. It is generally thought that the 
Active Galactic Nuclei (AGNs) or radio galaxies and quasars, the most distant discrete objects (at distances of giga parsecs and farther) seen in the universe,  
should trace the distribution of matter in the universe at that large scale and should therefore appear isotropically distributed 
from any observing position in the universe. But in recent years there have been many reports of large surveys, comprising sky-distribution data on millions of AGNs, showing large scale anisotropies that seem to be inconsistent with the CP \cite{36,37,7,26,34,8,19a,12a,10c,10d,Sie21,Se21,10e,10f,10g,10h}. The most glaring of these anisotropies are in the form of radio dipoles, which, though aligned with the CMBR dipole, differ in amplitudes by almost an order of magnitude. Bereft of these dipoles the radio source population may appear to possess an isotropic distribution \cite {12b}, nonetheless, the discrepancies observed in statistically significant (up to $\sim 10 \sigma$) dipole strengths may, perhaps, be pointers to a preferred direction in the sky \cite{7,10c,10d,Sie21,Se21,10e,10f,10g,10h}, something inconsistent with the CP. Effects of inhomogeneities on our understanding of cosmology can be found, for example, in \cite{ell09,ell11}.

However, if we ignore these and some other similar threats to CP and trust the assumption of homogeneity and isotropy 
{\em for the whole universe at all epochs}, then the line element can be expressed in the Robertson-Walker metric form \cite{9,6,4,5}.
\begin{eqnarray}
\nonumber
{\rm d}s^2=c^2 {\rm d}t^2 - R^2(t)\left[\frac{{\rm d}r^2}{\left(1-k\,r^2\right)^{1/2}}\right.\\
\label{eq:1}
\left.+r^2 {\rm d}\omega ^2\right],
\end{eqnarray}
where the only time dependent function is the scale factor $R(t)$. Here $r {\rm d}\omega =r {\sqrt{{\rm d}\theta ^2+\sin^2\theta {\rm d}\phi^2}}$ 
represents the angular line element. The constant $k$ is the curvature index that can take one 
of the three possible values $+1, 0$ or $-1$ and $(r, \theta, \phi)$ are the time-independent comoving coordinates.

Using Einstein's field equations, space curvature $k/R^2$ can be expressed in terms of the Hubble parameter $H$ and the density parameter $\Omega$ \cite{9,6,5}
\begin{eqnarray}
\label{eq:2}
\frac{k}{R^2}=(\Omega-1)\frac{H^2}{c^2},
\end{eqnarray}
where $\Omega=\Omega_{\rm m}+\Omega_{\rm r}+\Omega_\Lambda$ with $\Omega_{\rm m}$,  $\Omega_{\rm r}$, $\Omega_\Lambda$  as the matter density, radiation density and vacuum energy (dark energy) density parameters respectively. The space is thus flat ($k=0$) only if $\Omega=1$. The present value of a parameter will be denoted by a subscript $(_o)$.

In general it is not possible to express the comoving coordinate distance $rR_{\rm o}$ in terms of the cosmological redshift $z$ of the source 
in a close-form analytical expression and one may have to evaluate it numerically. For example, in the 
 $\Omega_\Lambda \neq 0, \Omega_{\rm o}=1$, matter-dominated world-models, $rR_{\rm o}$ is given by \cite{9}
\begin{eqnarray}
\label{eq:5}
rR_{\rm o}=\frac{c}{H_{\rm o}}\int^{z}_{0}\frac{{\rm d}z}{\left[\Omega_\Lambda+\Omega_{\rm m}(1+z)^3\right]^{1/2}}.
\end{eqnarray}
For a given finite $\Omega_\Lambda$, one can evaluate $rR_{\rm o}$ from Eq.~(\ref{eq:5}) by a numerical integration. 

However, for $\Omega_\Lambda = 0, \Omega_{\rm o}=\Omega_{\rm m}$ cosmologies, it is 
possible to express $rR_{\rm o}$ as an analytical function of redshift \cite{15,11}
\begin{eqnarray}
\label{eq:11}
rR_{\rm o}\!\!=\!\!\frac{cz\left[1+z+\sqrt{1+ z\Omega_{\rm o}}\right]}
{H_{\rm o}(1+z)\!\!\left[1+ z\Omega_{\rm o}/2+\sqrt{1+ z\Omega_{\rm o}}\right]}.
\end{eqnarray} 

From the expression~(\ref{eq:11}), one finds that as $z \rightarrow \infty$, $r R_{\rm o}$ converges to a finite value $2c/(H_{\rm o}\Omega_{\rm o})$, 
though the range of possible values of 
coordinate distance $rR_{\rm o}$ extends up to infinity. 
It turns out that {\em all} finite density FRW world-models, starting with a big bang, have a   
particle
horizon \cite{5}. 
It is thought that a finite horizon exists because there is only a finite amount of time since the big bang singularity 
(corresponding to $z \rightarrow \infty$), and that photons could have travelled only a finite distance within 
the finite age of the universe.  However, as we mentioned earlier,  for Milne's empty universe ($\Omega_{\rm o}= 0$), 
there is no finite horizon limit and the whole infinite universe is visible to any observer at any time (see also \cite{22}). 
Therefore the argument that a ``finite horizon'' arises in cosmological models because photons could have travelled only a finite distance 
since the big bang singularity, does not hold good in the most general case.
\begin{figure*}[t]
\begin{center}
\includegraphics[width=16cm]{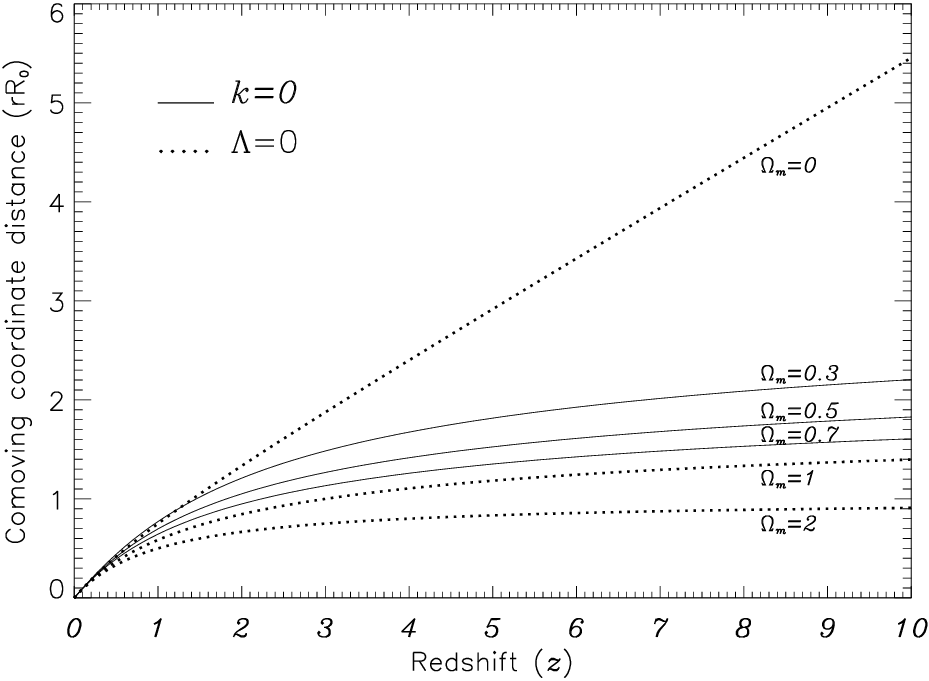}
\caption{A plot of the comoving coordinate distance $r R_{\rm o}$ (in units of $c/H_{\rm o}$) against  redshift for various world models with 
different matter density, $\Omega_{\rm m}$. The 
continuous lines depict flat space world models ($k=0$) while the dotted lines represent $\Lambda = 0$ cosmologies.}
\end{center}
\end{figure*}

Figure (2) show plots of comoving coordinate distance $r R_{\rm o}$ for various FRW world models. From Fig. 2 it is clear that the object 
horizon is at infinity for $\Omega_{\rm m}=0$ world models but it moves to smaller ($r R_{\rm o}$) values as $\Omega_{\rm m}$ increases.
In Table 1 we have listed the maximum values  
of the comoving coordinate distance $r R_{\rm o}$ (in units of $c/H_{\rm o}$) at the horizon ($z \rightarrow \infty$) for each model.
Appearance of particle
horizon in a world model is generally interpreted as that different parts of the universe in that model 
did not get sufficient time to interact with each other and thus may have yet no causal relations and therefore 
could not have achieved uniformity everywhere.  
Therefore inflation is invoked in which an exponential expansion of space takes place at time $t \sim 10^{-35}$ sec by a factor of  
$\sim 10^{28}$ or larger 
and the space-points now far apart (and thus apparently not in touch with each other so they appear to be causally unrelated) were actually 
much nearer before  $t \sim 10^{-35}$ sec or so and could have had time to interact with each other before inflation.

A crucial point that somehow seems to have been overlooked (or ignored) in these deliberations is that the question of horizon 
comes up only when CP, i.e. homogeneity and isotropy at all times, holds good to begin with as then only we could apply Robertson-Walker element with a time-dependence through a single scale parameter $R(t)$. 
And it is only then that horizon makes an appearance which in turn has given rise to  
the oft-discussed question of the uniformity and homogeneity of the universe at large. As long as we make use of the Robertson-Walker metric 
we are guaranteed that the universe was {\em ever} homogeneous and isotropic, and that a single parameter $R(t)$ can describe the past as well as determine the future of the universe. The Robertson-Walker line element ensues from 
the assumption of homogeneity and isotropy, and then starting from Eq.~(\ref{eq:1}) we are led to Eqs.~(\ref{eq:5}) or (\ref{eq:11}), 
where the presence of a horizon is inferred. However that in itself may  not 
imply a non-existence or lack of homogeneity as horizon itself makes an appearance in models where to begin with homogeneity is presumed. All we
find from calculations is that in such an isotropic and homogeneous universe, the light signals in a finite value of the the parameter time $t$ may  
not, in general, cover the whole available range of space coordinate $r$ in the universe. In fact in some of the world models, even in an 
infinite time, all $r$ do not get covered by light signals emitted from a point (``event horizon''). 
\begin{table}[t]
\begin{center}
\caption{Particle horizon for various FRW world models}
\vspace{2mm}
\begin{tabular}{@{}rrrrrrccc}
\hline
$k$ && $\Omega_\Lambda$ && $\Omega_{\rm m}$  
& &$rR_{\rm o}H_{\rm o}/c({z \rightarrow \infty})$ \\
\hline
\hline
$-1$ && 0 && 0 & & $\infty$ \\
0 && 0.7 && $0.3$ & & $3.2$\\
0 && 0.5 && $0.5$ & & $2.6$\\
0 && 0.3 && 0.7 & & $2.3$ \\
0 && 0 && 1 & & $2$\\
$+1$ && 0 && 2 & & $1$\\
\hline
\end{tabular}
\end{center}
\end{table}

Cause and effect seem to have been reversed in their roles  in this particular problem. 
It is not that because horizon exists so uniformity is not possible, ironically it is where a uniformity is present to 
begin with that we seem to end up with a horizon problem. In these models we assign only a single parameter $t$, and all other 
parameters describing the universe like the scale factor $R(t)$, density parameter $\Omega(t)$, Hubble parameter $H(t)$, deceleration parameter $q(t)$ 
etc. at any given time $t$ to be the same {\em everywhere} (even beyond object or event horizons wherever we might encounter such  
horizons). 
For instance, in a model where the density parameter $\Omega(t)$ is a function of only $t$ and is independent of spatial coordinates $r,\theta, \phi$, it implies that $\Omega(t)$ is uniform in spatial coordinates and it is in such cosmological models we find that light signals from any point cover only a finite distance in a finite time even though the possible range of $r$ may extend to distances beyond that. But that does not mean that it violates conditions of homogeneity and isotropy. It is the observed isotropy of CMBR which gives impetus to CP, which in turn is a basic ingredient in all cosmological models where horizon makes an appearance.  
It yet remains to be seen whether such horizons would still arise in models where one does not begin with CP and 
the consequential Robertson-Walker line element, and it is there one may have to deal with a genuine non-uniformity problem.

Actually if we follow the standard arguments in the literature then inflation in one sense makes the application of CP 
difficult. Even if it might alleviate the problem of object 
horizon, yet it gives rise to much more acute event-horizon problems. After all even just before inflation began, there would be particle
horizons, 
which because of a rapid expansion of the universe due to inflation will become even more ``remote'' from each other ending up in growth of large 
number of event horizons, with all such regions of the universe never able to interact with each other. 
Thus such a universe will comprise huge number of large patches still isolated from each others. Then how can one still apply the cosmological 
principle to such a disjointed universe which would conflict with our starting assumption of universal homogeneity and isotropy. We cannot then even use Robertson-Walker line element to describe the geometry of the whole universe 
and then all our conclusions about the cosmological 
models would have to be abandoned, with us back to square one. 

Once again, inflation or no inflation, 
these horizons are encountered only in the models where we have already assumed CP.
However, if we do want to really 
examine the question of homogeneity or its absence then we need to abandon the standard model based on the Robertson-Walker metric and then 
with some new model, where possibility of large scale anisotropy or inhomogeneity is assumed to begin with, and then one has to examine if in such models also 
we come across horizons and if so, then we have a genuine problem to contend with.
Examples of generalizations of the FRW class are indeed known \cite{ple06} with discussions on how to solve the horizon problem in an inhomogeneous universe \cite{cel98} or how isotropization of universe could occur in a class of exact anisotropic cosmological solutions of Einstein's equations \cite{wai98} etc.
\section{The Flatness Problem}
In the so-called flatness problem, the present density of the universe is observed to be very close (within an order of magnitude) to the critical density value needed for a zero curvature, i.e., $0.1<\Omega_{\rm o}<1$. Since the density departs rapidly from the critical 
value with time, the early universe must have had a density even closer to the critical density, so much so that if we extrapolate the density 
parameter to the epoch of inflation ($t \sim 10^{-35}$ sec) we find it to be near unity within an extremely small fraction of order 
$\sim 10^{-53}$. This leads to the question how the initial density came to be so closely 
fine-tuned to the critical value. To avoid this fine tuning, on one hand a standard argument prevalent in literature is that the universe {\em must be flat} ($k=0$). We shall show that such an argument is hardly of any essence, and the flatness problem, as it is posed, is not falsifiable. Alternatively, cosmic inflation has been proposed to resolve the fine-tuning issue along with the horizon problem \cite{2}
but as we will show inflation arguably creates perhaps more problems than it solves as far as the flatness problem is concerned.

Using $H=\dot{R}/R$, we can rewrite Eq.~(\ref{eq:2}) as
\begin{equation}
\label{eq:61.25a}
(\Omega-1) = \frac {k c^2}{H^2R^2} = \frac {k c^2}{\dot{R}^2}.
\end{equation}
For all world models with a big bang origin, ${R} \propto t^{1/2}$ near the big bang event (see e.g., \cite{3}), implying  $\dot{R}^2 \propto t^{-1}$ or $\Omega \rightarrow 1$ as $t \rightarrow 0$. This of course is the reason why in all such models, $(\Omega-1)$ can be extremely small in the early universe. 

Comparing the density parameter at an earlier epoch to the present value, say, for the open universe models ($k=-1$ and $\Omega<1$), we can write 
\begin{equation}
\label{eq:61.26}
(1-\Omega) =  \frac {\dot{R}^2_{\rm o}}{\dot{R}^2}(1-\Omega_{\rm o}),
\end{equation}
which can be written as
\begin{equation}
\label{eq:61.27}
(1-\Omega)=\epsilon \;(1-\Omega_{\rm o}),
\end{equation}
where $\epsilon$ could be an extremely small number depending upon the earlier epoch of reference. Now for a given world model and for the chosen epoch, $\epsilon$  may be a definite number, though we may know it only very approximately, perhaps only to an order of magnitude.  
For instance, for the epoch of inflation ($t \sim 10^{-35}$ sec), $\epsilon \sim 10^{-53}$, while for the Planck epoch ($t \sim 10^{-43}$ sec), $\epsilon \sim 10^{-61}$ \cite{1,9,10}.

To comprehend the consequences better, in Eq.~(\ref{eq:61.27}) we write $(1-\Omega)=\eta$ and $(1-\Omega_{\rm o})=\eta_{\rm o}$ to get
\begin{equation}
\label{eq:61.28}
    \eta=\epsilon \;\eta_{\rm o}.
\end{equation}
Here both $\eta$ and $\eta_{\rm o}$ lie between 0 and 1 for our open-universe model. That immediately implies that $\eta\le\epsilon$ and that $\eta$ cannot be larger than $\epsilon$ by even a tiniest amount. For instance, if we have $\eta=\epsilon(1+\epsilon)$, then 
$\eta_{\rm o}=\eta/\epsilon=(1+\epsilon)$. which violates the condition that $\eta_{\rm o}$ is between 0 and 1. 

 Now if the universe is flat ($k=0$) then inflation of course plays no part in this respect as it cannot make it any more flat. However in nearly flat universe scenario, inflationary theories purportedly alleviate the problem of fine-tuning by proposing that the universe in an interval of $\sim 10^{-32}$ seconds expands exponentially by a factor of $\sim 10^{28}$ in its linear size, thereby decreasing the curvature $k/R^2$ to a value close to zero and thereby bringing the density parameter of the universe very close to the required value of unity. However the huge expansion factor ($\sim 10^{28}$) in size then has to be extremely 
fine-tuned so that the resulting density parameter is such that $\eta$ does not exceed  $\epsilon \;(\sim 10^{-50})$ by even a tiniest amount. This assumption of inflation factor in a rather tight range, does it not imply replacing the erstwhile fine-tuning problem with another but more severe form of fine-tuning? 

The so-called fine-tuning in non-inflationary models is not really a fine-tuning as that is the nature of the FRW cosmological models and it depends upon the epoch chosen for the  investigation of the density parameter, but the fine-tuning implied in the inflationary models has to be just right at the end of the inflation. Does it really alleviate the fine-tuning problem in a fruitful manner. In fact if inflation brings the value of $\eta$ down by a large factor so as to match the present conditions, it would mean that before the inflationary era, for $\eta$ to be a moderate value ($\sim 1$),
from Eq.~(\ref{eq:61.25a}) the expansion rate need to be also a more moderate value ($\dot{R}\sim c$) near the big bang, a condition that could be a problem in the FRW models to satisfy, where $\dot{R} \rightarrow \infty$ as $t \rightarrow 0$. Does not the remedy seem to be worse than the malady, if any?

Further, this type of fine-tuning argument can be applied to almost any present value of the observed density of the universe. 
What is implied here is that even in a hypothetical, almost empty,  universe where the 
density of universe is say,  
$\rho_{\rm o} \sim 10^{-56}$ gm/cc or so (with density parameter $\Omega_{\rm o} \sim 10^{-28}$), having only a mass equivalent to that of Earth alone 
to fill the whole universe, from Eq.~(\ref{eq:61.27}) the density parameter at the epoch of inflation would differ from unity by the same 
fraction, of order $\sim 10^{-53}$. 
Is there really any substance in this type of arguments as even a mass equal to that of earth alone spread over the 
universe will lead to the same low departures from unity of $10^{-53}$? In fact even the 
presence of a mere single observer would imply the same departures from unity of $10^{-53}$. On the other hand if the whole universe were filled with water (proverbial deluge!) with density $\rho \sim 1$ gm/cc, equivalent to a density parameter $\Omega \sim 10^{28}$, one could still argue that the universe is flat otherwise it would require a fine tuning of the density parameter near the inflation era ($10^{-35}$) sec to be unity within a fraction of $10^{-25}$, an extremely small number.
Thus a use of fine tuning 
argument to promote $k = 0$ model amounts to {\em a priori} rejection of all models with $k \ne 0$, because the density parameter in all 
FRW world models gets arbitrarily close to unity as we approach the epoch of the big bang. That is the property of all 
these FRW models. That way, irrespective of the actually measured present-density value, 
we could use any sufficiently early epoch and use the ``extreme fine-tuning'' arguments to reject all non-flat models. But that is not what we would ever call any theory to be a falsifiable one. 

Without casting 
any whatsoever aspersions on the inflationary theories, we point out that one cannot use these type of arguments to support inflation. Further, it has been shown \cite{14} that for $\Omega_\Lambda\ne 0$ models, there exist non-flat FRW models for which $\Omega_{\rm o} \sim 1$ throughout the entire history of the universe, and that these really are not fine-tuned models. From an 
examination of the flatness problem quantitatively for all cosmological models it
has been concluded \cite{12} that the flatness problem does not exist, not only for the cosmological models
corresponding to the currently popular values of $\lambda$ and $\Omega_{\rm o}$ values but indeed for all FRW models with $\lambda \ne 0$.

As has been pointed out \cite{13}, from the type of arguments used in literature, one might 
consider a flat universe to be infinitely fine-tuned, since it assumes $\Omega_{\rm o}$ to be 
identically one, thereby making it the most unnatural choice.
By opting for a flat universe, the least probable out of three possible curvature values, we seem to be following the example of Copernican  
epicycles on philosophical grounds. 
Further, if $k = 0$, then 
inflation does not have a role to play here as it cannot flatten it further. And if $k \ne 0$ then inflation cannot make it $k = 0$, even 
though it might bring the density parameter closer to unity.
In fact by assuming a flat model we are assuming the ultimate finest-ever tuning imaginable where even the least amount of perturbation on this unstable equilibrium 
model (in the form of an excess or deficiency of the smallest amount of matter from the critical density
- a single particle or atom extra or missing!) can ultimately take the universe away from the flat-space model to a curved one. 

There are recent reports of a possible evidence for a closed universe, using Planck lensing observations \cite{Ha19,Va20}. Although it might appear inconsistent with other Planck information, but it still raises the question that we should always be putting fundamental cosmological assumptions, viz. cosmological principle and flatness, to observational tests instead of assuming them a priori.

\section{Conclusions}
We demonstrated that the horizon problem arises in FRW models where homogeneity and isotropy of the universe---\`a la cosmological principle--- is assumed to begin with for all epochs. Therefore, to look for extraneous reasons for the observed isotropy of the universe within the framework of these models is not very meaningful. We do not yet know if horizon problem would arise in non-FRW models. We also showed that the flatness problem, as posed in literature, is not even falsifiable and that a fine-tuning does not discriminate between various world models and a use of fine tuning 
argument amounts to  {\em a priori} rejection of all models with $k \ne 0$. Thus one has to rely only upon observations and not on such argument to justify if the universe is really flat or not. Further, to resolve both horizon and flatness problems, inflation is not a must, though other compelling grounds, viz. prediction of an almost scale invariant spectrum of primordial scalar perturbations confirmed by CMBR observations, might be there to justify its occurrence. In the inflationary scenario (as well as other alternatives like the bouncing or the loitering universe models), one assumes a 3+1 geometry, employing the scale factor with only a time dependence. 
However, this is permissible only if to begin with itself one presumes the validity of the cosmological principle, i.e., homogeneity and isotropy everywhere and for all times. Therefore it is redundant to attempt to smoothen out any large scale inhomogeneities, presumed to be there because of the presence of a horizon that one encounters in this already homogeneous universe. In fact, horizon or no horizon, there is no need for any further smoothening by invoking inflation, or some other alternative, for a universe which is already smooth on large scales.

\end{document}